\documentclass[prb,twocolumn,floatfix]{revtex4}
\usepackage{graphicx}
\usepackage{dcolumn}
\usepackage{bm}

\newcommand{\vdc}{\ensuremath{V_{\text{dc}}}}
\newcommand{\vrf}{\ensuremath{v_{\text{rf}}}}
\newcommand{\vset}{\ensuremath{v_{\text{SET}}}}
\newcommand{\lqo}{\ensuremath{q_{0}}}
\newcommand{\fo}{\ensuremath{f_{0}}}

\newcommand{\zin}{\ensuremath{Z_{\text{in}}}}

\newcommand{\rn}{\ensuremath{R_{n}}}
\newcommand{\rd}{\ensuremath{R_{d}}}
\newcommand{\gd}{\ensuremath{G_{d}}}
\newcommand{\pr}{\ensuremath{P_{r}}}

\newcommand{\ec}{\ensuremath{E_{c}}}

\newcommand{\ehz}{\ensuremath{e/\sqrt{\mathrm{Hz}}}}
\newcommand{\aehz}[1]{\ensuremath{#1\:\ehz}}
\newcommand{\e}[1]{\ensuremath{\times 10^{#1}}}
\newcommand{\units}[1]{\ensuremath{\mathrm{#1}}}
\newcommand{\amount}[2]{\ensuremath{#1\:\units{#2}}}
\newcommand{\cp}{\ensuremath{C_{p}}} 
\newcommand{\csig}{\ensuremath{C_{\Sigma}}}

\newcommand{\vg}{\ensuremath{V_{g}}}

\newcommand{\iv}{$I$-$V$}

\begin{document}

\title{On-Chip Matching Networks for Radio-Frequency Single-Electron-Transistors}

\author{W. W. Xue}\email{weiwei.xue@dartmouth.edu}
\affiliation{6127 Wilder Lab, Department of Physics and Astronomy,
Dartmouth College Hanover, NH, 03755 USA}
\author{Z. Ji}
\affiliation{Department of Physics and Astronomy, Rice University, Houston, TX 77005 USA}
\author{B. Davis}
\affiliation{6127 Wilder Lab, Department of Physics and Astronomy,
Dartmouth College Hanover, NH, 03755 USA}
\author{Feng Pan}
\affiliation{6127 Wilder Lab, Department of Physics and Astronomy,
Dartmouth College Hanover, NH, 03755 USA}
\author{J. Stettenheim}
\affiliation{6127 Wilder Lab, Department of Physics and Astronomy,
Dartmouth College Hanover, NH, 03755 USA}
\author{T. J. Gilheart}
\affiliation{6127 Wilder Lab, Department of Physics and Astronomy,
Dartmouth College Hanover, NH, 03755 USA}
\author{A. J. Rimberg}\email{ajrimberg@dartmouth.edu}
\affiliation{6127 Wilder Lab, Department of Physics and Astronomy,
Dartmouth College Hanover, NH, 03755 USA}

\begin{abstract}
In this letter, we describe operation of a radio-frequency superconducting single electron transistor (RF-SSET) with an on-chip superconducting $LC$ matching network consisting of a spiral inductor $L$ and its capacitance to ground \cp. The  superconducting network has a lower \cp\ and gives a better matching for the RF-SSET than does a commercial chip inductor. Moreover, the superconducting network has negligibly low dissipation, leading to sensitive response to changes in the RF-SSET impedance. The charge sensitivity $\delta q=\aehz{2.4\e{-6}}$ in the sub-gap region and energy sensitivity $\delta \varepsilon=1.9\hbar$ indicate that the RF-SSET is operating in the vicinity of the shot noise limit.
\end{abstract}

\maketitle

\begin{figure}
\includegraphics[width=6.5cm]{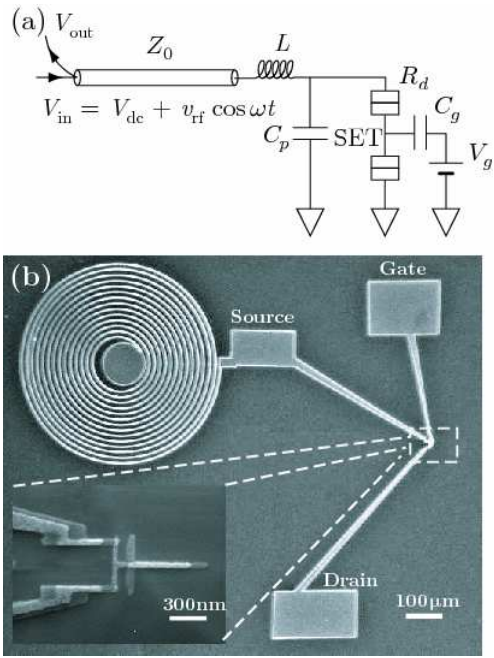}
\caption{\label{fig1} (a) Idealized model of an $LC$ matching network. (b) Optical micrograph of an on-chip matching network prior to wire bonding.  The apparent inductor linewidth is set by the resolution of the image. The inset shows an electron micrograph of the SET\@.}
\end{figure}

With growing interest in quantum computation,\cite{Shor:1994,Grover:1997} spin-based qubits,\cite{Elzerman:2004a,Petta:2005} the quantum properties of nanomechanical resonators,\cite{Knobel:2003,LaHaye:2004} and quantum measurement\cite{Braginsky:1992,Devoret:2000} much attention has been focused on ultra-fast charge detectors such as the radio-frequency single electron transistor (RF-SET).\cite{Schoelkopf:1998,Aassime:2001,Aassime:2001a,Brenning:2006} In rf mode, the SET is embedded in an $LC$ network as illustrated in Fig.~\ref{fig1}(a) allowing a working bandwidth of tens of MHz and avoiding $1/f$ noise from amplifiers and background charges.  The $LC$ network usually consists of a commercial chip inductor $L$ and its parasitic capacitance to ground \cp; such networks, however, have drawbacks such as losses and relatively large \cp\ that degrade the performance of the SET\@. In this letter, we describe RF-SSETs with on-chip fully superconducting $LC$ matching networks. Although our best charge sensitivity $\delta q =\aehz{2.4\e{-6}}$ does not quite match the record to date\cite{Brenning:2006}, our SET  and matching network design are not yet fully optimized. Furthermore, our measurement is in the sub-gap region for which transport occurs via a combination of Cooper pair and quasiparticle tunneling. The backaction of the SET, the rate at which it dephases a measurement,\cite{Clerk:2002,Makhlin:2001a} is predicted to be smaller in the sub-gap region than in the above-gap region for which Coulomb blockade of quasiparticles dominates.\cite{Aassime:2001,Aassime:2001a,Brenning:2006}

Fig.~\ref{fig1}(a) shows an idealized model for an on-chip superconducting matching network. One end of the SET is connected to an Al spiral inductor $L$, which is then connected via a coaxial cable to room temperature electronics. The other end of the SET is grounded directly to the cable shield. The inductor $L$, the SET differential resistance \rd, and the stray capacitance \cp\ from the inductor and SET bonding pads to ground form an $LCR$ network with resonant frequency $\fo\approx\frac{1}{2\pi\sqrt{L\cp}}$. A carrier wave with frequency \fo\ and rms amplitude \vrf\ is applied to the network and the reflected signal is measured. The reflection coefficient at resonance is given by  $\Gamma=\frac{\zin-50}{\zin+50}$ where the input impedance of the network $\zin=\frac{L}{\rd\cp}$.   In order to optimize the charge sensitivity, \zin\ should be impedance matched to the \amount{50}{\Omega} coaxial cable at the point of maximal change in SET conductance with charge.  The unloaded quality factor $Q =\sqrt{\frac{L}{\cp}}\frac{1}{Z_{0}}$ determines the amplitude of the rf signal applied to the SET $\vset=2Q\vrf$ and the resonance bandwidth $\fo/Q$.

An on-chip superconducting matching network has three advantages in comparison with a commercial chip inductor.  First, because \cp\ is smaller for an on-chip network, better impedance matching can be attained at higher frequencies, resulting in a larger resonance bandwidth for a given $Q$.  Second, an on-chip network can be extended to multi-pole matching networks\cite{Abrie:2000} that can further increase the bandwidth, possibly allowing measurements on nanosecond time scales. Finally, our on-chip $LC$ networks are entirely superconducting at our measurement temperature and have negligible loss at radio frequencies.  In comparison, the input impedance at resonance of a matching network that includes normal metals has loss terms arising from dissipation in the inductor $L$ or capacitor \cp\  in addition to the transformed SET impedance $\frac{L}{\rd\cp}$. The reflection coefficient for a lossy network is therefore less sensitive to changes in \rd. While fully on-chip matching networks have been used previously, they have generally included some normal metal components.\cite{Stevenson:2002,LaHaye:2004}

 Fig.~\ref{fig1}(b) shows an optical micrograph of an on-chip network. The network is fabricated together with the SET by e-beam lithography and double-angle shadow evaporation of Al. The number and spacing of the turns of the spiral inductor (linewidth \amount{3}{\mu m}, line spacing \amount{20}{\mu m}) determines $L$.  The inset of Fig. 1(b)  shows a scanning electron micrograph of the SET with junction area about 40$\times$60 $\rm nm^{2}$. The center of the spiral inductor is wire bonded using an Al wire to the central pin of a coaxial cable and the ground lead of the SET is similarly bonded to the cable shield. 
 
\begin{figure}
\includegraphics[width=7cm]{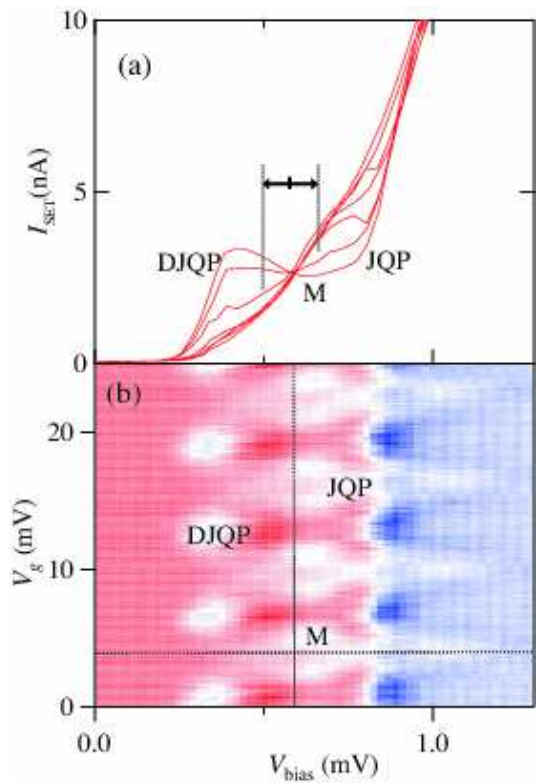}
\caption{\label{fig2}(a) \iv\ curves of sample A for various $V_{g}$. The modulation at the DJQP and JQP features is about \amount{2}{nA}. Point M shows the dc bias \vdc\ for optimal RF-SET operation and the arrows and vertical dashed lines show the peak to peak rf amplitude at the SET  $\vset=2Q\vrf\approx\amount{160}{\mu V}$. (b) False-color image of \gd\ versus \vdc\ and \vg.  Experimentally determined optimal values of \vdc\ and \vg\ for rf operation are indicated by the dashed lines.}
\end{figure} 
 
 The measurements were carried out in a $^3$He refrigerator at the base temperature of \amount{290}{mK}. Copper-stainless steel powder filters in the cryostat and $\pi$-type filters at room temperature were used to eliminate high frequency noise. A low-noise HEMT amplifier and directional coupler were located in the cryostat at a temperature of around \amount{2.8}{K}. We made two samples with the same SET design and similar total normal-state resistance   \rn: sample A was coupled to a 12-turn spiral inductor and sample B to a 14-turn inductor.  DC \iv\ curves were measured with custom-made low noise current and voltage amplifiers, and the SET differential conductance $\gd=1/\rd$ via standard lock-in techniques. Results for sample A are shown in Fig.~\ref{fig2}.  Features associated with two sub-gap charge transport cycles, the Josephson-quasiparticle (JQP) and double Josephson-quasiparticle (DJQP) cycles are clearly visible; for a detailed discussion see Ref.\ \onlinecite{Thalakulam:2004} and references therein.   We determined the SET charging energy $\ec=e^{2}/2\csig=\amount{205}{\mu eV}$ where \csig\ is the total SET capacitance from the location of the DJQP feature and $\rn=\amount{25}{k\Omega}$ from the slope of the \iv\ curve at high bias.   Similar measurements for sample B gave $\ec=\amount{222}{\mu eV}$ and $\rn=\amount{26}{k\Omega}$.

\begin{figure}
\includegraphics[width=6.5cm]{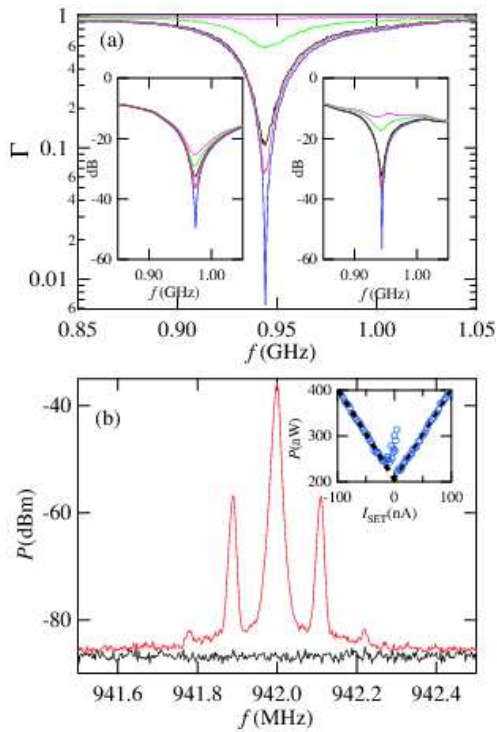}
\caption{\label{fig3}(a)  $\Gamma$ versus frequency for sample B for different \rd\ as determined from lockin measurements of \gd. Top to bottom: center of the gap (pink), $\rd=\amount{40}{k\Omega}$ (green), \amount{28.2}{k\Omega} (black), \amount{22.2}{k\Omega} (red), \amount{19.2}{k\Omega} (blue). The inset shows reflected power \pr\ versus frequency for the low impedance SET and Panasonic chip inductor (left) and for sample B (right). Left inset, top to bottom: center of the gap (pink), $\rd=\amount{37}{k\Omega}$ (green), \amount{21}{k\Omega} (black), \amount{17.8}{k\Omega} (red), \amount{12}{k\Omega} (blue).  Right inset, top to bottom: center of the gap (pink), $\rd=\amount{40}{k\Omega}$ (green), \amount{28.2}{k\Omega} (black), \amount{22.2}{k\Omega} (red), \amount{19.2}{k\Omega} (blue). (b)  Power spectrum of the RF-SSET output for  a \amount{110}{kHz} $0.01e$~rms excitation. The lower line is the noise floor with no rf signal applied on the SET. Inset: total system noise at \fo\ versus the dc SET current in the absence of applied rf power.}
\end{figure}

  We found that sample B (14 turn inductor) was better matched to the coaxial line, with near perfect matching at $\rd=\amount{20}{k\Omega}$. With the SET biased near the center of the gap ($\rd\gg\amount{1}{M \Omega}$), virtually all the input signal is expected to be reflected. This expectation is in agreement with the data for sample B in the right inset of the Fig.~\ref{fig3}, which shows the power \pr\ reflected by the tank circuit for different \rd. The top curve, which indicates \pr\ for  $\rd\gg\amount{1}{M \Omega}$, has no dip in \pr\ at resonance, only a background slope due to details of the rf setup. The reflection coefficient $\Gamma$ shown in Fig.~\ref{fig3} is obtained from the data in the right inset.  We assume $\Gamma=1$ for the largest \rd\ and, using the top curve as a reference, calculate $\Gamma$ at different \rd\ by taking the difference between the other curves and the reference. Virtually identical results are obtained by fitting a line to the background slope of the top curve and using the fit as the reference instead.
  
 With decreasing \rd, $\Gamma$ decreases over two orders of magnitude at resonance, reaching a minimum of $\Gamma=0.006$ at $\rd=\amount{19.2}{k\Omega}$.  Our bandwidth of roughly $\amount{50}{MHz}\sim\fo/Q$ is roughly three times larger than that for measurements with similar $Q$ but lower resonant frequency.\cite{Aassime:2001,Aassime:2001a} From the ratio of  $\Gamma$ at resonance for any two different \rd, and the expressions for $\Gamma$, \zin\ and \fo\ given above,  we calculate $Q\approx20$, $L\approx\amount{170}{nH}$ and $\cp\approx\amount{0.17}{pF}$ for which the optimal $\rd=\amount{20}{k\Omega}$. To compare with a commercial inductor, we fabricated a low-impedance sample with $\rn=\amount{13}{k\Omega}$ and coupled it to the coaxial cable through a Panasonic ELJ \amount{82}{nH} chip inductor.  \pr\ for this sample, for which $\fo=\amount{975}{MHz}$ and the optimal $\rd=\amount{11}{k\Omega}$, is shown in the left inset of Fig.~\ref{fig3}.  We estimate $\cp\approx\amount{0.34}{pF}$; the relatively large value of \cp\ prevents matching to larger \rd.   Furthermore, a dip in \pr\ of about \amount{11}{dB} appears at resonance for very large \rd, indicating that only about 8\% of the input power is reflected. The other 92\% is lost to dissipative processes in the matching network.\cite{Roschier:2004}
  
  Response of our RF-SSETs to a charge excitation was measured with a spectrum analyzer; typical data are shown in Fig.~\ref{fig3}(b).  The charge sensitivity $\delta q$ is determined from the rms charge excitation amplitude \lqo\ and the signal-to-noise ratio in dB (SNR) of a sideband from $\delta q = (\lqo/\sqrt{2\,\text{BW}})10^{-\text{SNR}/20}$ where BW is the measurement bandwidth.  The best charge sensitivity for sample B is $\delta q=\aehz{2.4\e{-6}}$, about three times better than that achieved with a lossy $LC$ network and the same rf setup.\cite{Thalakulam:2004} We calibrated our system noise temperature for sample B by measuring the total noise power versus the dc SET current $I$ (Fig.~\ref{fig3}(b), inset).  At higher bias, the noise varies linearly with $I$ due to the SET shot noise.\cite{Aassime:2001a}  The contribution from the HEMT amplifier is determined by the crossing point of the two fitting curves at $I=0$. We obtain an amplifier noise power $P_{n}=\amount{210}{aW}$ for a measurement bandwidth $\text{BW}=\amount{3}{MHz}$, giving a noise temperature $T_{n}=\frac{P_{n}}{\text{BW}k_{B}}=\amount{5.3}{K}$.  The uncoupled energy sensitivity of sample B is $\delta \varepsilon=\frac{(\delta q)^{2}}{2C_\mathrm{SET}}= 1.9\hbar$, approaching the shot noise limit for the RF-SET.\cite{Korotkov:1999,Devoret:2000} Without the contribution from the cryogenic amplifier we estimate $\delta \varepsilon=1.2\hbar$. For sample A, we measured similar values of $\delta q=3.1\times 10^{-6} e/\sqrt{Hz}$ and $\delta \varepsilon=3.1\hbar$.

  Embedding the RF-SSET in the on-chip matching network shows potential for studying the shot noise of the SET for either rf or dc biases by making several improvements in our system. First, the \rd\ for optimal charge sensitivity in sample B was about \amount{35}{k\Omega} (point M in Fig.~\ref{fig2}(a)), while near-perfect matching occurred at $\rd =\amount{20}{k\Omega}$.  Further improvements in the matching network design should allow us to reduce \cp\ and increase $L$ for better matching with higher \rd\ without lowering the resonant frequency. Also, using a HEMT amplifier with lower noise temperature will improve both the charge and uncoupled energy sensitivity. Finally, improved fabrication techniques for the SET may also lead to a better charge sensitivity.
  
This work was supported by the NSF under Grant No.\ DMR-0454914 and by the ARO under Agreement No.\ W911NF-06-1-0312.

\end{document}